\documentstyle[psfig,twocolumn,prl,aps]{revtex}

\begin{document}

\draft

\title{Concentration of Charge Carriers and Anomalous Gap
Parameter in the Normal State of High-$T_c$ Superconductors} 

\author{Andrew Das Arulsamy$^{a,b,}$\thanks{E-mail: sadwerdna@hotmail.com}}

\address{$^a$Department of Physics, \
Center for Superconducting and Magnetic Materials, \
Faculty of Science, \
National University of Singapore, \
2 Science Drive 3, \
117542 Singapore, Singapore \\} 

\address{$^b$No. 22, Jalan Melur 14, \
Taman Melur, 68000 Ampang, \
Selangor DE, \
Malaysia \\} 

\maketitle

\begin{abstract}
Fermi-Dirac statistics has been utilized by introducing the average ionization energy ($E_I$) as an additional anomalous energy gap in order to derive the two-dimensional concentration of charge carriers and the phenomenological resistivity model for the superconducting polycrystalline materials. The best fitted values of $E_I$ and the charge carriers' concentration ranges in the vicinity of 4 to 9 meV and 10$^{16}$ m$^{-2}$ respectively for the superconducting single crystal samples and polycrystalline compounds synthesized with various compositions via solid-state reactions.  The phenomenological resistivity model is further redefined here based on the gapless nature of charge-carriers' dynamics within the Cu-O$_2$ planes that corresponds to anomalous Fermi liquid behavior, which is in accordance with the nested Fermi liquid theory.  
\end{abstract}


\section{Introduction}

Since the reported discovery of high-$T_c$ superconductors (HTSC) based on the copper oxide materials, the scientific communities have been swept with revolutionary excitement.  This leads to intensified focus on this particular material both in term of theoretical and experimental research throughout the world.  The discovered Ba-La-Cu-O system by Bednorz and Muller\cite{bednorz1} with a $T_c$ of 33 K have had led to subsequent findings of superconductivity in La-Sr-Cu-O at 40 K\cite{cava2,tarascon3} and later in Y-Ba-Cu-O system at 90 K\cite{wu4}. Actually the La-Ca,Sr,Ba-Cu system were first synthesized and studied by Raveau {\it et al.}\cite{michel5}.  
The conduction mechanisms in the normal state of HTSC such as the out-of-plane and in-plane resistivities are intriguing due to the highly anisotropic behavior.  Anisotropic normal state electrical properties such as metallic-like conduction in the Cu-O$_2$ planes and semiconducting-like conduction between those planes\cite{anderson6} are believed to be one of the central concerns for applications\cite{ito7}.  This in turn may lead one to extract some clues to understand the mechanism of high-$T_c$ superconductivity\cite{ito7,fisher8}. Interestingly, Levin {\it et al.}\cite{levin9} using the $T$-dependence on anisotropy $\rho_c/\rho_{ab}$ showed the conductivity relationship between $\sigma_c$ and $\sigma_{ab}$ as: $\sigma_{ab}$ = $(a + bT^{-2/3})\sigma_c$. This $T^{-2/3}$ dependence point towards a 2-dimensional (2D) localization and hopping.  Quite recently, Basov {\it et al.}\cite{basov10} have analyzed the interlayer infrared conductivity of HTSC.  They found that discrepancy occurred between superfluid density, $\rho_s$ and difference between normal state ($N_n$) and superconducting state ($N_s$) spectral weights i.e. $\rho_s$ $>$ $N_n$ $-$ $N_s$. This discrepancy was only revealed for $c$-axis electrodynamics but could not find a similar conflict in $ab$-plane ($\rho_s$ = $N_n$ $-$ $N_s$), which can be interpreted as a gapless behavior of charge carriers in $ab$-planes.   
PrBa$_2$Cu$_3$O$_7$ compound gives a unique non-superconducting, semiconducting-like normal state electrical property and the nature of this unique property was discussed intensively in terms of hole doping in Cu-O$_2$ planes\cite{almasan11,zhao12,das13}. As a matter of fact, semiconducting-like behavior of $\rho(T)$ corresponds to localization of charge carriers connected with oxygen disorder in Cu-O$_x$ chains and the disorder in Pr$^{3+}$, Pr$^{4+}$ distribution\cite{romanenko14}.  Furthermore, positron annihilation studies\cite{hoffmann15,shukla16} showed that PrBa$_2$Cu$_3$O$_7$ compound has the Fermi surface due to Cu-O$_x$ chains' state.  The concept of charge concentration in high-$T_c$ cuprates is very ambiguous\cite{kubo17}.  However, it is well established that $T_c$ of cuprates depends on holes' concentration within Cu-O$_2$ planes where carrier concentration is determined by counting the valence number\cite{kubo17,muroi18}. Very recently, concentration of interlayer electrons and its resistivity model for HTSC was derived via Fermi-Dirac statistics with an additional gap constraint, which is the average ionization energy ($E_I$)\cite{arulsamy19}.  In this letter however, a phenomenological polycrystalline resistivity model is derived based on $E_I$ as well and is redefined here to obtain an $ab$-plane resistivity model by incorporating gapless nature of charge carriers in Cu-O$_2$ planes. Besides that, estimation for concentration of charge carriers and $E_I$ via best fitting in the normal state of HTSC are also illustrated in this work.  Apart from that, estimation of resistivity transition such as from metallic to semiconductor or vice versa upon substitution in crystalline 123 phase Pr$_x$Y$_{1-x}$Ba$_2$Cu$_3$O$_7$ system is also highlighted.  Ions' average ionization energies were calculated from ref.\cite{emsley20}.

\section{Charge Carriers' Concentration and Resistivity Models}

Tunneling electron's distribution has been derived via Fermi-Dirac (FD) statistics using average ionization energy ($E_I$) as an anomalous constraint, and is given below\cite{arulsamy19}

\begin{eqnarray}
f_e(E) & = & exp[-\mu-\lambda(E+E_I)].
\label{eq:1}
\end{eqnarray}

Similarly, probability function for holes can be derived and is given in Eq.~\ref{eq:2} below, where it is obtained by inserting the restrictive condition $E_{hole}$ $=$ $E_{initial state}$ $-$ $E_I$.  $E_{initial state}$ is the energy of a particle in a given system at a certain initial state and ranges from 0 to -$\infty$.  This can be justified as follows; for a hole to occupy a lower state $M$ from initial state $N$ is more probable than to occupy state $L$ if the condition $E_I$($M$) $<$ $E_I$($L$) at certain temperature, $T$ is satisfied.  

\begin{eqnarray}
f_h(E) & = & exp[\mu+\lambda(E-E_I)].
\label{eq:2}
\end{eqnarray}
						                                           
Note that Eq.~\ref{eq:1} and Eq.~\ref{eq:2} can be reduced to the standard Fermi-Dirac distributions if $E_I$ approaches 0.  The importance of average $E_I$'s inclusion is that, it can be interpreted as a $c$-axis gap, which represents the gap that tied to a neutral atom and could be varied with ionic substitutions. The respective normal state concentration of electrons\cite{arulsamy19} and holes are given by

\begin{eqnarray}
n & = & \int_0^\infty{f_e(E)N_e(E)dE},
\label{eq:3}
\end{eqnarray}

\begin{eqnarray}
p & = & \int_{-\infty}^0{f_h(E)N_h(E)dE}.
\label{eq:4}
\end{eqnarray}
	
$\mu$ $=$ $-E_F/k_BT$, $\lambda$ $=$ $1/k_BT$ are carefully chosen to confine the total number of particles in a given system equals to $n$.  $N_e(E)$ and $N_h(E)$ are 2D density of states (DOS) that can be derived from the Schrodinger equation. $N_e(E)$ and $N_h(E)$ are derived as $m^*_e/\pi\hbar^2$ and $m^*_h/\pi\hbar^2$ respectively and note that both 2D DOS are independent of $E$. Solutions of integral Eq.~\ref{eq:3} and Eq.~\ref{eq:4} give the concentration of tunneling electrons between Cu-O$_2$ layers, $n$ and concentration of holes, $p$ in the normal state of HTSC as 

\begin{eqnarray}
n & = & \frac{m_e^*k_BT}{\pi\hbar^2}exp\left[\frac{E_F - E_I}{k_BT}\right].
\label{eq:5}
\end{eqnarray}

\begin{eqnarray}
p & = & \frac{m_h^*k_BT}{\pi\hbar^2}exp\left[\frac{-E_F - E_I}{k_BT}\right].
\label{eq:6}
\end{eqnarray}

Eq.~\ref{eq:5} and Eq.~\ref{eq:6} are expressed in term of ionization energy, where $k_B$ is Boltzmann constant, $m^*_e$ and $m^*_h$ are electrons' and holes' effective masses respectively, $\hbar$ $=$ $h/2\pi$, $h$ is Plank constant and $E_F$ is Fermi level.  Moreover, from Eq.~\ref{eq:5} and Eq.~\ref{eq:6}, concentration of charge carriers in the normal state of polycrystalline HTSC can be written in the form of geometric mean which is given by 

\begin{eqnarray}
\sqrt{np} & = & \frac{k_BT}{\pi\hbar^2}\sqrt{m_e^*m_h^*}exp\left[\frac{-E_I}{k_BT}\right].
\label{eq:7}
\end{eqnarray}

This equation is free from Fermi level, hence concentration of charge carriers in superconducting polycrystals can be calculated by first estimating $E_I$, which is described in the following paragraph. Scattering among fermions leads to the conclusion of $1/\tau$ $\propto$ $T^2$\cite{anderson6} and has been confirmed empirically that $\rho$ $\propto$ $T^2$ as well\cite{maeno21,hashimoto22,yoshida23}. Furthermore, by utilizing metallic resistivity equation of $\rho$ $=$ $m/ne^2\tau$ where $e$ and $\tau$ are electron's charge and mean scattering free time respectively\cite{cyrot24} then one can derive the normal state phenomenological model for HTSC as given in Eq.~\ref{eq:8} below.  This were carried out by substituting $m$ = $m^*_e$ $=$ $m^*_h$, $n$ $=$ $\sqrt{np}$, and $1/\tau$ $=$ $AT^2$ into metallic resistivity equation. 

\begin{eqnarray}
\rho_{poly}(T,E_I) & = & A\frac{\pi\hbar^2}{e^2k_B}Texp\left[\frac{E_I}{k_BT}\right].
\label{eq:8}
\end{eqnarray}

$A$ is independent of temperature. Eq.~\ref{eq:8} above can be used in the normal state resistivity of polycrystalline superconducting samples to evaluate the variations in $E_I$.  Substituting this $E_I$ into Eq.~\ref{eq:7} at certain temperature, $T$ leads to the estimation of charge carriers' concentration. Subsequently, an $ab$-plane resistivity model can be derived quite naturally by taking anomalous gap = 0 due to gapless behavior of charge carriers in Cu-O$_2$ planes\cite{basov10}, thus Eq.~\ref{eq:8} can be simply rewritten as 

\begin{eqnarray}
\rho_{ab}(T) & = & A\frac{\pi\hbar^2}{e^2k_B}T.
\label{eq:9}
\end{eqnarray}

This latter phenomenon could be due to change in the scattering rate that results from a different Fermi surface phase space argument in accordance with the nested Fermi liquid theory\cite{virosztek25}.  

\section{Discussion}

Polycrystalline resistivities, $\rho_{poly}(T)$ of crystalline YBa$_2$Cu$_3$O$_7$\cite{hagen26} samples (B and C) were obtained using 

\begin{eqnarray}
\rho_{poly}(T) &=& \sqrt{\rho_c(T)\rho_{ab}(T)}. \label{eq:10}
\end{eqnarray}

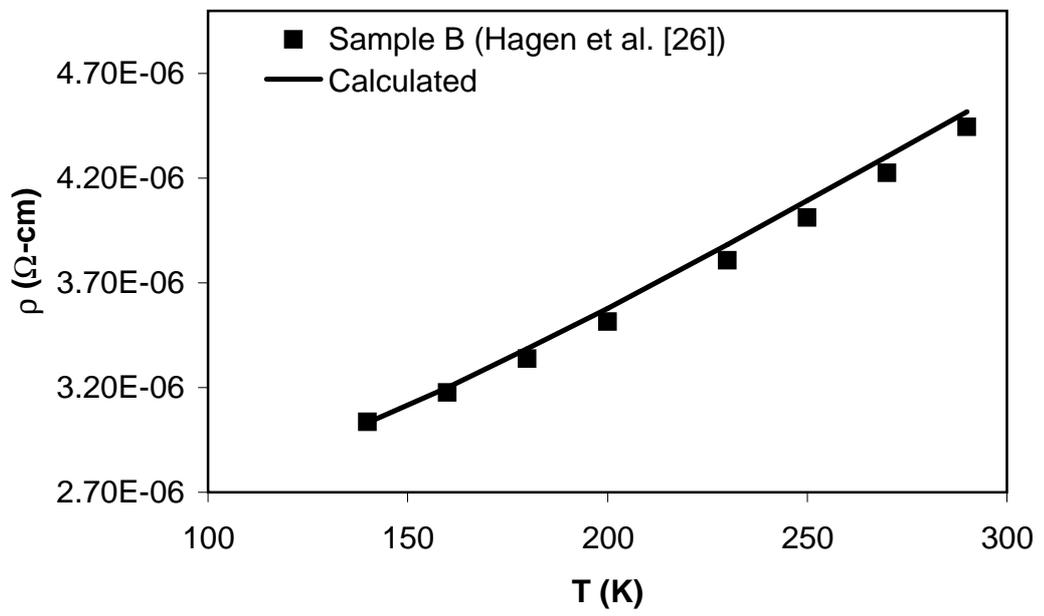
\begin{figure}

\caption {The resistivity curves of crystalline YBa$_2$Cu$_3$O$_7$ (Sample B: with volume of unit cell = 630 $\times$ 640 $\times$ 75 $\mu$m) of 123 phase were calculated using Eq.~\ref{eq:10}. Both $\rho_c(T)$ and $\rho_{ab}(T)$ were computed using Montgomery's algorithm. The single crystals (A and B) were grown by a flux method based on BaO. The ratio $a/c$ for sample B is approximately 8.}
\label{fig:1}
\end{figure}

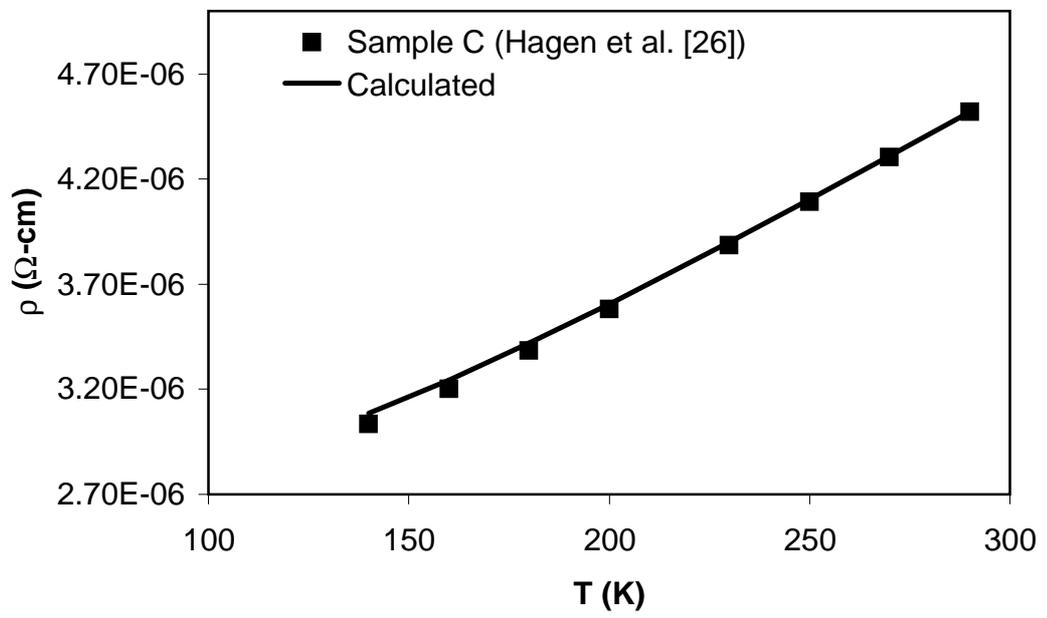
\begin{figure}

\caption {The resistivity of the crystalline YBa$_2$Cu$_3$O$_7$ (Sample C: with volume of unit cell = 390 $\times$ 400 $\times$ 25 $\mu$m) of 123 phase were obtained by utilizing Eq.~\ref{eq:10}. Both $\rho_c(T)$ and $\rho_{ab}(T)$ were computed using Montgomery's algorithm. The ratio $a/c$ for sample C is approximately 16.}
\label{fig:2}
\end{figure}

It is assumed that Eq.~\ref{eq:10} is valid to convert $\rho_c(T)$ and $\rho_{ab}(T)$ of single crystals to $\rho_{poly}(T)$ in order to analyze the effect or changes of $A$ parameter between single crystals and polycrystals. Consequently, one can study the scattering effect solely on $A$ parameter. Fitting of experimental $\rho_{poly}(T)$ with Eq.~\ref{eq:8} gives respective values for anomalous gap and concentration of charge carriers in the order of 4-9 meV and $10^{16}$ m$^{-2}$ as listed in Table~\ref{table:1}. These fittings are depicted in Fig.~\ref{fig:1} and Fig.~\ref{fig:2} for samples B and C respectively.  Furthermore, resistivity curves of polycrystalline Ga1212\cite{adachi27}, Y$_{1-x}$Pr$_x$SrBaCu$_3$O$_7$\cite{das13}, YBa$_2$Cu$_3$O$_{7-x}$\cite{dou28,kilcoyne29} and Tl1212\cite{lee30} samples are utilized by calculating the slope, $d\rho(T)$/$dT$ in the normal state of respective samples.  This slope would enable one to determine $E_I$ via Eq.~\ref{eq:8} and followed by determination of concentration of charge carriers in the normal state at 300 K and 150 K, which are also listed in Table 1. The above concentration of charge carriers was estimated by substituting the best fitted $E_I$ into Eq.~\ref{eq:7}. All samples have charge carriers' concentration in the order of 10$^{16}$ m$^{-2}$, however, parameters $A$ and $E_I$ vary accordingly with samples. Note that $A$ is equals to $A\pi\hbar^2/e^2k_B$ for simplicity's sake.  It is quite interesting to note that all Y123 samples including the single crystals (sample A and B) have values for $E_I$ between 7.69 and 8.25 meV.  Contrary to that, $A$ parameter is approximately 100 times lower for single crystals (10$^{-8}$) than polycrystals (10$^{-6}$) where this could be due to the nature of scattering in crystalline samples. Note also that the magnitude of $\rho_{poly}(T)$ estimated from Eq.~\ref{eq:10} for single crystals are lower as mentioned previously. Similarly, $A$ and $E_I$ parameters for 1212 phase compounds, Tl(Sr$_{1.6}$Sm$_{0.4}$)CaCu$_2$O$_{7-x}$, Tl(Sr$_{1.6}$Dy$_{0.4}$)CaCu$_2$O$_{7-x}$\cite{lee30} and GaSr$_2$(Y$_{0.6}$Ca$_{0.4}$)Cu$_2$O$_7$\cite{adachi27} are in the order of 10$^{-5}$ and 5-6 meV respectively.  Substitution effect in Y$_{1-x}$Pr$_x$SrBaCu$_3$O$_7$ system is found to be not in favor of superconductivity where, Pr substitution increases anomalous gap, $E_I$ and $A$ parameter. Parallel to this, comparison between polycrystalline samples, YBa$_2$Cu$_3$O$_{7-x}$ (Sample A)\cite{kilcoyne29} and Y(Ba$_{0.85}$Pb$_{0.15}$)$_2$Cu$_3$O$_{7-x}$ (Sample B)\cite{kilcoyne29} expose a reduction in $E_I$ that suggests Pb substitution into Ba sites is somewhat favorable.  Further comparisons between samples prepared by different listed researchers are not discussed here since the exact synthesis method varies from one researcher to another.   
In another context, $\rho_c(T)$ transition that occurs in superconducting Y$_{1-x}$Pr$_x$Ba$_2$Cu$_3$O$_7$ system is in fact due to the differences in average $E_I$ of Pr$^{3+,4+}$ and Y$^{3+}$. The valence number, $z+$ ($=$ 3+) and $q+$ (= 3+) correspond to ions, Pr$^{z+}$ and Y$^{q+}$ respectively.  Since $E_I$ of Y$^{3+}$ (1260 kJmol$^{-1}$) $>$ $E_I$ of Pr$^{3+}$ (1210 kJmol$^{-1}$) then substituting Y with Pr will reduce $E_I$ $-$ $E_F$ which is the $c$-axis gap parameter\cite{arulsamy19} for single crystals thus, reduces $\rho_c(T)$. Similar substitutions would also decrease the $E_I$ gap parameter and $\rho_{poly}(T)$ which is for polycrystals that can be verified from Eq.~\ref{eq:8}. In contrast, the experimental data obtained by Jiang {\it et al.} (single crystals)\cite{jiang31} and Das {\it et al.} (polycrystals)\cite{das13} expose a contradicting relation where $\rho_c(T)$ and $\rho_{poly}(T)$ increases with Pr substitutions into Y sites. Thus valency of Pr is expected to be greater than 3+ due to Pr$^{4+}$ ions' contribution\cite{klenscar32}. Parallel to this, Meng {\it et al.}\cite{meng33} have employed PVL theory to determine valency of Pr in superconducting state.  They have found that valency of Pr is vital in producing a superconducting Y$_{1-x}$Pr$_x$Ba$_2$Cu$_3$O$_7$ compound. In their calculations, valency of Pr is $<$ 3.15 to get a superconductor compound for all Pr doping but a valency of $>$ 3.15 gives a superconducting state for only up to a limited Pr doping.  Therefore, using the concept presented in section II, the minimum value of Pr's valence state, which is obviously above 3+ that contribute to non-metallic behavior, can be calculated from linear algebraic equation given below 

\begin{eqnarray}
\frac{\delta}{j}\sum^{z+j}_{i=z+1}{E_{Ii}} + \frac{1}{z}\sum^{z}_{i=1}{E_{Ii}} & = & \frac{1}{q}\sum^{q}_{i=1}{E_{Ii}}
\label{eq:11}
\end{eqnarray}

The first term, $\frac{\delta}{j}\sum^{z+j}_{i=z+1}{E_{Ii}}$ in Eq.~\ref{eq:11} above has $i$ $=$ $z$ + 1, $z$ + 2,..., $z$ + $j$ and $j$ $=$ 1, 2, 3,.... It is solely due to Pr$^{4+}$ ions contribution or caused by reaction of the form Pr$^{3+}$ $-$ electron $\to$ Pr$^{4+}$, hence $j$ is equals to 1 in this case and $\delta$ represents the additional contribution from Pr$^{4+}$. The second ($i$ $=$ 1, 2, 3, ..., $z$) and last ($i$ $=$ 1, 2, 3, ..., $q$) terms respectively are due to reaction of the form Pr $-$ 3(electrons) $\to$ Pr$^{3+}$ and Y $-$ 3(electrons) $\to$ Y$^{3+}$. Recall that $q$ = $z$ = 3+ and $i$ = 1, 2, 3,... represent the first, second, third, ... ionization energies while $j$ = 1, 2, 3, ... represent the fourth, fifth, sixth, ... ionization energies. Therefore, $z$ + $\delta$ gives the minimum valence number for Pr which is calculated to be 3.0133 from Eq.~\ref{eq:11}. This critical value corresponds specifically for $\rho(T)$ variations in the normal state behavior where $\rho(T)$ decreases with doping if and only if $z$ + $\delta$ $<$ 3.0133 and vice versa if $z$ + $\delta$  $>$ 3.0133.  Those changes in $\rho(T)$ will eventually affect $d\rho(T)/dT$ slope transition in term of metallic-like or semiconducting-like normal state properties as mentioned for pure 1212 phase polycrystals\cite{arulsamy19}.  
In order to justify this critical value, $\rho_c(T)$ measurements of Y$_{1-x}$Pr$_x$Ba$_2$Cu$_3$O$_7$ single crystals\cite{jiang31} that indicate a transition of $d\rho_c(T)/dT$ slope from positive to negative with $x$ doping were used. Thus, from the above-presented approximation, $z$ + $\delta$ or valence state of Pr for $x$ = 0.13 to 0.42 samples is $>$ 3.0133 since there is an increment of $\rho_c(T)$ with Pr doping. In contrast, $z$ + $\delta$ value should be above 3.15 as mentioned previously in accordance with Meng {\it et al.}\cite{meng33}. Although 3.15 is larger than what have been anticipated (3.0133), but one should note that 3.15 corresponds only to superconducting behavior without considering the normal state behavior. Therefore, by considering only the normal state behavior, decrement of $\rho_c(T)$ with increasing Pr doping only occurs if $z$ + $\delta$ $<$ 3.0133. In short, if 3.0133 $<$ $z$ + $\delta$ $<$ 3.15, then the materials are superconductors for all $x$ doping but $d\rho_c(T)/dT$ changes from positive to negative with $x$. However, if $z$ + $\delta$ $<$ 3.0133 then the materials are also superconductors for all doping, $x$ but $d\rho_c(T)/dT$ remains positive and $\rho_c(T)$ will decrease with $x$. Finally, if $z$ + $\delta$ $>$ 3.15, then the materials are superconductors for only upto a certain $x$ doping in which, $d\rho_c(T)/dT$ changes from positive to negative and $\rho_c(T)$ would increase with $x$. Note that the prediction above is only valid for very pure materials without any significant impurity phases. $\rho_c(T)$ data at high temperatures\cite{jiang31} surprisingly decreases with further Pr doping from $x$ = 0.42 to 0.55 though it remains semiconducting. This phenomenon could be possibly due to variation in the magnitude of $z$ + $\delta$ where the value for $z$ + $\delta$ might change from; $z$ + $\delta$ (at $x$ = 0.42) $>$ $z$ + $\delta$ (at $x$ = 0.53) $>$ $z$ + $\delta$ (at $x$ = 0.55). I.e., Pr$^{z+\delta}$ varies with doping or concentration, $x$. Apart from that, as evidenced in experimental data, the phenomenological $ab$-plane resistivity model (Eq.~\ref{eq:9}) indicates a linear relationship between temperature and in-plane resistivity due to gapless nature of charge carriers, thus, pointing towards an anomalous Fermi liquid behavior within Cu-O$_2$ planes. 

\section{Conclusions}

Two-dimensional Fermion characteristics have been employed to derive charge carriers' concentration.  Anomalous gap parameter ($E_I$) and charge carriers concentration that has been estimated via best fitting for all superconducting samples of polycrystalline and crystalline are in the order of 4-9 meV and 10$^{16}$ m$^{-2}$ respectively.  As a matter of fact, variations that occurred in $\sqrt{np}$, $A$ and $E_I$ are due to variations in sample's purity, methods of preparation and its compositions.  Phenomenological $ab$-plane resistivity model (Eq.~\ref{eq:9}) indicates a linear relationship between temperature and in-plane resistivity due to gapless nature of charge carriers, signifying an anomalous Fermi liquid behavior in accordance with nested Fermi liquid theory. In addition, disorder of valence state in certain elements such as Pr$^{3+}$, Pr$^{4+}$ in term of average $E_I$ also contributes to the normal state behavior such as from metallic-like to semiconducting-like behaviors. In other words, $d\rho_c(T)/dT$'s varies from positive to negative or vice versa with selected doping or substitutions at certain temperature. Simply put, polycrystalline resistivity's slope is dependent on average ionization energy, where valence state of Pr which is above 3+ could be estimated from $d\rho_c(T)/dT$ transition upon substitution of Pr into Y sites as Eq.~\ref{eq:11} states. 

\section*{ACKNOWLEDGMENT}

ADA would like to thank the National University of Singapore and Physics department for the financial assistance. The author is also grateful to A. Innasimuthu, I. Sebastiammal, A. Das Anthony and Cecily Arokiam for their partial financial assitance.

\begin{table}

\caption {Calculated values of the temperature independent parameter ($A$), anomalous gap parameter ($E_I$), resistivity slope ($d\rho(T)/dT$) and the concentration of charge carriers ($\sqrt{np}$) at 300 K and 150 K in the normal state of high-$T_c$ superconducting compounds. Note that the magnitudes of $E_I$ are in the range of 4-9 meV and the charge carriers' concentration is determined to be in the order of 10$^{16}$ via appropriate fittings. $d\rho(T)/dT$ is calculated from the experimental $\rho_{poly}(T)$ plots. Note that $\sqrt{np}$ is obtained from Eq.~\ref{eq:7}. $k_B$ is the Boltzmann constant.} 
\label{table:1}
\end{table}

\end{document}